\documentclass{emulateapj}
\usepackage{apjfonts}
\usepackage{amsbsy}

\newcommand{\up}[1]{{\rm #1}}

\newcommand{\vv}{\upsilon}
\newcommand{\kms}{{\rm km\, s}^{-1}}

\newcommand{\msun}{M_{\odot}}
\newcommand{\hmsun}{{h^{-1}\msun}}

\newcommand{\MH}{M_\up{h}}
\newcommand{\RH}{R_\up{h}}
\newcommand{\Mres}{M_\up{res}}

\newcommand{\MBH}{M_\up{BH}}

\renewcommand{\MBH}{M_\bullet}

\newcommand{\bdv}[1]{\pmb{#1}}

\newcommand{\beeq}{\vspace{10pt}\begin{equation}}
\newcommand{\eneq}{\vspace{10pt}\end{equation}}

\begin{document}

\title{The Most Massive Black Holes in the Universe:\\
Effects of Mergers in Massive Galaxy Clusters}

\author{Jaiyul Yoo$^1$, Jordi Miralda-Escud\'e$^{1,2}$, David H. Weinberg$^1$, 
Zheng Zheng$^{3,4}$, and Christopher W. Morgan$^{1,5}$}

\altaffiltext{1}{Department of Astronomy, The Ohio State University, 
140 West 18th Avenue, Columbus, OH 43210; 
jaiyul, dhw, morgan@astronomy.ohio-state.edu}

\altaffiltext{2}{Institute de Ci\`encies de l'Espai (IEEC-CSIC)/ICREA, Spain;
 miralda@ieec.uab.es}

\altaffiltext{3}{School of Natural Sciences, Institute for Advanced Study,
Einstein Drive, Princeton, NJ 08540; zhengz@ias.edu}

\altaffiltext{4}{Hubble Fellow}

\altaffiltext{5}{Department of Physics, United States Naval Academy, 572C
Holloway Road, Annapolis, MD 21402}

\slugcomment{accepted for publication in The Astrophysical Journal}
\shorttitle{THE MOST MASSIVE BLACK HOLES IN THE UNIVERSE}
\shortauthors{YOO ET AL.}

\begin{abstract}
Recent observations support the idea that nuclear black holes grew by gas
accretion while shining as luminous quasars at high redshift, and they
establish a relation of the black hole mass with the host galaxy's spheroidal
stellar system.  We develop an analytic model to calculate the expected
impact of mergers on the masses of black holes in massive clusters of galaxies.
We use the extended Press-Schechter formalism to generate Monte Carlo merger
histories of halos with a mass $10^{15}\hmsun$.
We assume that the black hole mass function at $z=2$ is similar
to that inferred from observations at $z=0$ (since quasar activity declines
markedly at $z<2$), and we assign black holes to the progenitor halos
assuming a monotonic relation between halo mass and black hole mass.
We follow the dynamical evolution of subhalos within larger halos,
allowing for tidal stripping, the loss of orbital energy by dynamical
friction, and random orbital perturbations in gravitational encounters
with subhalos, and we assume that most black holes will efficiently merge
after their host galaxies (represented as the nuclei of
subhalos in our model) merge.
Our analytic model reproduces numerical estimates of the subhalo mass
function. We find that mergers can increase the mass of the most massive
black holes in massive clusters typically by a factor $\sim$ 2,
after gas accretion has stopped.
In our ten realizations of $10^{15}\hmsun$ clusters,
the highest initial ($z=2$) black hole masses
are $5-7 \times 10^9\msun$, but four of the clusters contain black holes
in the range $1-1.5\times 10^{10}\msun$ at $z=0$.  Satellite galaxies may
host black holes whose mass is comparable to, or even greater than, that of the
central galaxy.  Thus, black hole
mergers can significantly extend the very high
end of the black hole mass function.

\end{abstract}


\section{Introduction}
\label{sec:int}

  The idea that quasars are powered by accretion of gas in a disk
orbiting a massive black hole \citep{salp,zeld,lynden} requires that
enough black holes should exist at the present time to account for all
the mass that had to be accreted over the past history of the universe
by all observed quasars \citep{solt}. Much progress has been made over
the last decade by discovering black holes in galactic nuclei, measuring
their masses and characterizing the population (\citealt{mago}). Using
the surprisingly tight relation that is observed between the black hole
mass and the velocity dispersion of the spheroidal stellar population of
the host galaxy (\citealt{laura,karl,mbh} and references therein), one
finds that the average mass density of the population of nuclear black
holes is roughly equal to what is required for the amount of light that
has been emitted by all Active Galactic Nuclei (AGN), if the radiative
efficiency is close to 10\%, a typical value expected for geometrically
thin accretion disks (see, e.g., \citealt{barger,yutre,franc}). This
implies (with the caveat of the substantial observational uncertainties
that are still present in the determination of the black hole mass
density and the integrated AGN emission) that a major contribution to
the growth of the black hole mass density is the accretion of gas that
is responsible for AGN, and that contributions from radiatively
inefficient accretion (which may arise in super-Eddington or strongly
sub-Eddington accretion flows) are at most comparable to the accretion
from geometrically thin disks.

Black hole mergers do not change the total mass density in the black hole
population, so they do not alter this argument, but they can alter the mass
distribution of black holes. Observationally, we have relatively little handle 
on the black hole mass function outside the range $10^7-10^9\msun$ that 
dominates the average mass density. The
most luminous quasars of which we have evidence at high redshift imply
black hole masses close to $10^{10} \msun$ (e.g., \citealt{fan2}),
assuming they are radiating at the Eddington luminosity. The most
massive black hole discovered so far is in M87, with a mass of
$\MBH\sim 3\times 10^9 \msun$. It is natural to assume that the highest
mass black holes may be lurking in some of the most massive elliptical
galaxies, but we do not know whether the power-law $\MBH-\sigma$
correlation extends up to these largest black hole masses. Similar
uncertainties plague the determination of the abundance of low-mass
black holes.

  A number of interesting questions can be raised in relation to
theoretical expectations for the most massive black holes. Under the
assumption that the Eddington luminosity cannot be exceeded (a point
recently questioned by Begelman, Volonteri \& Rees~[2006]), what is the
abundance of the most massive black holes we should expect? Do the cold
dark matter models for structure formation predict that the most
massive black holes
should generally reside in central cluster galaxies, or can they
sometimes reside in a massive galaxy that has only recently merged in a
cluster, still moving in the cluster outskirts?
How much could the most massive black holes have
grown by mergers? If we search today for black holes in the central
cluster galaxies of the most massive clusters, what is the black hole
mass we should expect to measure?

  With these questions in mind, in this paper we develop a dynamical
model for the evolution of the supermassive black hole population.
We use this model to examine the effect of black hole mergers on
the high-mass end of the present-day mass distribution.
The model we
develop aims to treat carefully the dynamics of a halo containing a
central black hole once it becomes an orbiting satellite within a larger
halo. The problem of the merger of two central black holes after their
halos are merged will not be considered here, and we simply obtain the
maximum merger rate by assuming that two black holes will always merge
on a short time scale once they become bound to each other
in the center of two merging halos. 
After the host halo of a black hole's parent galaxy
merges into a bigger halo, it becomes an
orbiting subhalo. The subhalo may either remain in orbit
indefinitely (in which case the black hole will not merge), or it
may be dragged by dynamical friction
to the center of the larger halo within which it is
orbiting and be tidally disrupted. In the latter case, a black hole
merger may ensue. We will present the result of a model where black
holes are initially distributed among host halos after they have been
formed by gas accretion, and then their mergers are followed according
to the dynamical evolution of their subhalos. Our model is similar to
that of \citet{marta}, though we treat the dynamical
evolution of the subhalos in greater detail. We shall focus in this
paper on the black holes that are present in a halo of $10^{15}\hmsun$
at the present time, representing a massive galaxy cluster.

Our implementation of a merger tree for dark matter halos is described
in \S~\ref{sec:pro}, and the method to follow the dynamical evolution
of merged subhalos is explained in \S~\ref{sec:dyn} and \S~\ref{sec:num}.
In \S~\ref{sec:res}, 
we specify the
initial conditions by which we populate halos with black holes of
different masses at an initial redshift. The results on the
importance of mergers for black holes in clusters, and the masses of
the central black holes, are presented in \S~\ref{sec:impact} and discussed in
\S~\ref{sec:con}.

\section{Properties of Dark Matter Halos}
\label{sec:pro}

  We develop in this paper a dynamical model of substructure in dark
matter halos. The first step is to construct a merger tree, starting
from a present day halo (which for the application explored in this
paper will represent a halo like that of the Coma cluster) and
going backwards in time
to generate all the merger events. This section describes how the
merger tree is generated, how each halo in the tree is assigned a
mass and a radius, and the choice we make for the halo density profiles.
In \S~\ref{sec:dyn}, we describe the second step, where we follow the dynamical
evolution of the halos after they have become satellites of a larger
halo, starting at the earliest time when the first low-mass halos form,
and moving forward to the present time.\footnote{Throughout the paper, 
we interchangeably use $satellite$, $subhalo$, and 
$substructure$ to refer to a distinct gravitationally self-bound halo in a 
larger dark matter halo.} 

\subsection{Merger History}
\label{sec:mh}

  We use the extended Press-Schechter formalism to generate the merger tree
\citep{ps,bcek}. This formalism is very
simple to use and is known to provide an adequate description for a
merger history of a halo of mass comparable to the Milky Way galaxy
(e.g., \citealt{bul1}), although in detail there are
deviations of the halo mass function from the results of
numerical simulations (e.g., \citealt{eke,st,jen}).

  We define $n(M,z)$ as the number density of halos of mass greater than
$M$ at redshift $z$, and $\sigma(M,z)$ as the extrapolated linear rms
fluctuation in spheres containing an average mass $M$.
The halo number density is related to
the fraction of the mass that is in halos of mass $M$ per unit
$\ln\sigma(M)$ by
\beeq
f(\sigma,z) = {M\over\bar\rho_m}{d n(M,z)\over d\ln\sigma},
\eneq
where $\bar\rho_m$ is the mean matter density of the universe. The
Press-Schechter model assumes that
\beeq
f=\sqrt{2\over\pi}{\delta_c\over\sigma}\exp
\left(-{\delta_c^2\over2\sigma^2}\right),
\eneq
where $\delta_c(z)$ is the critical threshold on the linear overdensity
for collapse at $z$. We adopt the approximate fitting formula for
$\delta_c(z)$ of Eke, Cole, \& Frenk (1996; see also, \citealt{car}) for a
$\Lambda$CDM universe. The idea for the merger tree is based on the
conditional probability that a particle in a halo of mass $M_2$ at
redshift $z_2$ was part of a halo of mass $M_1$ at redshift $z_1(>z_2)$ 
\citep{lac,col}.
\vspace{10pt}\begin{eqnarray}
f_{12}(M_1,M_2)d M_1&=&{1\over\sqrt{2\pi}}{{(\delta_{c1}-\delta_{c2})}
\over{(\sigma^2_1-\sigma^2_2)^{3/2}}} \nonumber \\
&&\times \exp\left[-{{(\delta_{c1}-\delta_{c2})^2}\over
{2(\sigma^2_1-\sigma^2_2)}}\right]
{{d\sigma^2_1}\over{d M_1}}d M_1,
\end{eqnarray}
where each subscript indicates the corresponding redshift. Then, the
mean number of halos of mass $M_1$ that merge into a halo of mass
$M_2$ over a time step $dt_1$ is $d N/d M_1=(d f_{12}/d t_1)(M_2/M_1)\,
dt_1$. The merger tree with mass resolution $\Mres$ is built by
generating new progenitors of mass $\Mres< M_1 < M_2/2$ with total
probability $P$ given by \citep{col}
\beeq
P=\int_{\Mres}^{M_2/2}{{d N}\over{d M_1}}\, d M_1~,
\label{eq:frag}
\eneq
and adding a rate of smooth mass accretion, $F$, equal to
\beeq
F=\int_{0}^{\Mres}{{d N}\over{d M_1}}{M_1\over M_2}\,d M_1 ~,
\eneq
to account for the mass of unresolved halos with $M_1 < \Mres$ that
are not discretely generated.

\subsection{Density Profile}
  Density profiles of halos from numerical simulations have been
fitted to a variety of forms \citep{nfw,moo2,fuk,sersic}. Here we
will be interested in describing profiles of satellite halos with
finite mass after they have merged into a larger halo. The profile of
\citet{nfw} has a density slope $d\log\rho/d\log r$
approaching 3 at large radius, giving rise to a logarithmic divergence
of the total mass, which demands the introduction of a cutoff radius.
We adopt instead the simple form of the Jaffe sphere:
\beeq
\rho_\up{h}(r)=\left({\MH\over{4\pi \RH^3}}\right)
{\RH^4\over{r^2(r+\RH)^2}} ~,
\label{eq:jaffe}
\eneq
where the Jaffe radius $\RH$ is defined as the radius inside which the
mass is half of the total halo mass $\MH$. 
This profile is steeper than that found in
purely collisionless numerical simulations
in the central part, but it
has the advantage of describing a finite mass halo with a very simple
form for the potential. In addition, black holes should in reality be
in the center of a galaxy, and the baryonic component will steepen the
profile near the center making it closer to that of a Jaffe sphere.

  In later sections, we will be using the one-dimensional velocity
dispersion of the Jaffe sphere at each radius for isotropic orbits,
which is found from the spherical Jeans equation \citep{bin} to be
\vspace{10pt}
\begin{eqnarray}
\label{eq:sig}
\sigma^2(r)&=&-{6G\MH\over \RH}\left({r\over \RH}\right)^2
\left(1+{r\over \RH}\right)^2\ln\left({r\over{r+\RH}}\right) \\
&&-{G\MH\over2\RH}\left[12\left({r\over \RH}\right)^3+18\left({r\over \RH}
\right)^2+4\left({r\over \RH}\right)-1\right]. \nonumber
\end{eqnarray}
\vspace{10pt}

\subsection{Radii of Dark Matter Halos}

  The evolution of a halo of mass $\MH$ after it merges with another
object depends not only on its mass, but on its half-mass radius $\RH$
as well, because in general denser objects will be less vulnerable to
tidal disruption and will more easily survive as orbiting satellites.
It is therefore important to estimate the radius of a halo as it
evolves, first by acquiring new mass through mergers and accretion, and
then by losing mass through tidal disruption once it has been
incorporated into a larger halo as a satellite.

  When a halo first appears in the merger tree at the resolution
mass $\Mres$,
we assign an initial radius to the halo, assuming the standard collapse
model in which a spherical top-hat overdense region turns around from
the Hubble expansion at radius $R_\up{t}$, and then virializes
at a collapse radius $R_\up{col}=R_\up{t}/2$.
We use the fitting formula of Bryan \& Norman (1998) for the mean
density of the non-linear collapsed object,
$\rho_\up{col} = \rho_\up{crit}(z) \Delta_\up{col}$,
where $\rho_\up{crit}$ is the critical density
at the redshift
$z$ of collapse, the mean matter density is $\rho_m(z) = \Omega_m(z)
\rho_\up{crit}$, and a flat cosmological model is assumed:
\beeq
\Delta_\up{col}=18\pi^2+82x-39x^2, ~~~ x=\Omega_m(z)-1 ~.
\eneq
The collapse radius is obtained from
\beeq
\MH={4\pi\over 3}\, R_\up{col}^3\rho_\up{col} ~.
\eneq
For the Jaffe density profile of halos we will be assuming in
this paper, the Jaffe radius $\RH$ that corresponds to this collapse
radius is obtained by equating the initial potential energy of a
homogeneous sphere at the turnaround radius $R_\up{t} = 2R_\up{col}$
to half of the potential energy of the Jaffe sphere, which gives
\beeq
\RH={5\over 6}\, R_\up{col}={5\over12}\,R_\up{t} ~.
\label{eq:rh}
\eneq

  However, as halos continue to merge, their radius should depend on the
detailed merger history. Halos found at a redshift $z$ in which most of
the mass has been recently accreted should have a mean density of the
order of $\rho_\up{col}$, but halos that acquired most of their mass at an
epoch much earlier than redshift $z$ and have since then accreted at a
low rate should have a much higher density, reflecting the density of
the universe at the time they were formed (e.g., \citealt{nfw}). To
simulate this process, the radii of halos in the merger tree are
evolved as follows: At any time step in which two halos of masses
$M_\up{h1}$ and $M_\up{h2}$ merge, and in addition an unresolved mass
$M_\up{acc}$
is accreted, we consider that the new halo of mass $\MH = M_\up{h1}+M_\up{h2}
+M_\up{acc}$ has formed from a configuration where the two halos and the
smoothed mass $M_\up{acc}$ turned around from a radius
$R_\up{t} = 2R_\up{col}$,
with $R_\up{col}$ given by equation (\ref{eq:rh}).
The total energy of the final halo with a Jaffe sphere profile after
virialization is half its potential energy, or $-G\MH^2/(4\RH)$, where
$\RH$ is the Jaffe radius of the final halo, and we equate this to the
total energy of the system at turnaround:
\vspace{10pt}
\begin{eqnarray}
-{GM^2_\up{h}\over 4\RH}&=&-{GM^2_\up{h1}\over 4R_\up{h1}}-
{GM^2_\up{h2}\over 4R_\up{h2}} \\
&&-{3\over 5}{G(M_\up{acc}^2+2M_\up{h1}M_\up{h2}+2M_\up{h1}M_\up{acc}+
2M_\up{h2}M_\up{acc})\over
2R_\up{col}} ~. \nonumber
\label{eq:merge}
\end{eqnarray}
In general, the gravitational potential energy of a homogeneous sphere
of mass $\MH$ and radius $2R_\up{col}$ is $(3/5)G\MH^2/(2R_\up{col})$.
In the last term of
equation~(\ref{eq:merge}), we have included the self-gravity of mass
$M_\up{acc}$, and the interaction terms of mass $M_\up{h1}$ with $M_\up{h2}$,
of mass $M_\up{h1}$ with $M_\up{acc}$, and of mass $M_\up{h2}$
and $M_\up{acc}$. The
self-gravity terms of $M_\up{h1}$ and $M_\up{h2}$ are not included
because these are already part of the total energy of the two merging
halos that were previously virialized.
In the case of a time step in which there is only smooth accretion,
we simply put $M_\up{h2}=0$ in equation~(\ref{eq:merge}).

\section{Substructure Dynamics}
\label{sec:dyn}

  This section describes the dynamical evolution of a dark matter halo
once it merges into a larger halo within which it orbits as a satellite.
At every point in the merger tree when two halos merge, the smallest of
the two halos becomes a satellite, and the large one is the host halo
within which the satellite orbits. The mass of a halo increases at every
step due to accretion and mergers until the moment it merges with
a halo larger than itself. After this time, the mass of the halo as a
satellite can only decrease due to tidal stripping.

  The model we use to compute the dynamical evolution is based on
ideas similar to those in the models
of \citet{tb1,tb2}, \citet{bul1,bul2}, and 
\citet{zb}. The major differences are the following:
(1) We compute the dynamical evolution of all the halos identified in
the merger tree, instead of following only the history of the most 
massive halo that eventually becomes the main halo at the present day.
(2) We continue to follow the evolution of satellites after their
host halo becomes a satellite itself.
(3) We include the random variations of orbits due to
relaxation of the satellites with each other.
This type of analytic model for the evolution of substructure has been
shown to be in reasonable agreement with $N$-body simulation results
\citep{tor,tb1,tb2,taf,zb}.

  When a halo merges and becomes a satellite, it is assigned an initial
orbit in its host halo. The evolution of the orbit and
the halo structure is then modeled including the effects of dynamical
friction, adiabatic orbital variations as the host halo grows, mass
loss from the tides of the host halo at the orbital pericenter, and
random orbital variations due to the gravitational deflections from
other satellites. Satellites are also assigned a core radius; when tides
have disrupted the satellite down to the core radius, the satellite is
considered to be completely dissolved. 
For the application we are
interested in this paper, the core radius of satellites
is set to the black
hole zone of influence whenever a black hole is present at the 
center of a satellite halo
(the way halos are populated with black holes will be specified
in \S~\ref{sec:res}): $R_\up{z}\equiv G\MBH/2\sigma_0^2$,
where $\sigma_0$ is the velocity dispersion of the Jaffe sphere at
$r=0$ in the absence of a black hole and $\MBH$ is the mass of a black hole
at the center.
We consider a satellite to be completely destroyed and merged
with its host halo when $\MH\leq\MBH$, and we assume that
their central black holes also merge at this point.
If the satellite does not contain a black hole, its
core radius is set to 1\% of the Jaffe radius.
We describe these dynamical treatments
in detail in the following subsections.

\subsection{Orbital Initial Conditions}
\label{sec:ioc}

  After a merger, a halo will initially move on an orbit with typical
radius $R_\up{col}$, the radius of collapse of the host halo of mass $\MH$
at the redshift of the merger introduced in \S 2.3.
Therefore, we place the halo on an initial orbit with the same energy as
a circular orbit at radius $R_\up{col}$:
\beeq
\epsilon_0 = { \vv_c^2 \over 2} -{G\MH\over \RH} ~
\ln\left(1+{\RH\over R_\up{col}} \right) \, ,
\eneq
where we have used the potential of a Jaffe sphere with half-mass radius
$\RH$, and
\beeq
\vv_c^2 (R_\up{col}) = {G\MH\over \RH + R_\up{col} } ~.
\label{eq:vcjaffe}
\eneq

  In addition to the orbital energy, each halo is also assigned an
initial orbital angular momentum. For this, we assume an isotropic
velocity distribution, i.e., we assume that the phase-space density
depends only on energy. This implies that the radial distribution of
particles in orbits of energy $\epsilon_0$, in the potential of the
Jaffe sphere we have adopted, is
\beeq
P(r)\, dr\propto 4\pi r^2d r\int d^3\vv \delta(\epsilon-\epsilon_0)\propto
r^2 dr \sqrt{\ln\left({r_\up{max}\over r}{{r+\RH}\over{r_\up{max}+\RH}}\right)}
~,
\label{eq:prob}
\eneq
where $r_\up{max}$ is the maximum radius at which the satellite can be located
for the given initial energy $\epsilon_0$.
To choose the initial orbital angular momentum of a halo after a
merger, we first generate a random radius $r$ after having calculated
the initial orbital energy $\epsilon_0$ with the distribution of
equation (\ref{eq:prob}), and then we compute the potential energy at
radius $r$ and the modulus of the velocity vector at this radius
required to have the orbital energy $\epsilon_0$. The velocity is
then assigned a random direction, which yields the orbital angular
momentum, and the corresponding pericenter and apocenter of the
orbit.

\subsection{Dynamical Friction}

  We compute the rate at which the orbital radius of a satellite
decreases owing to dynamical friction using the usual Chandrasekhar
formula. A satellite of mass $M_s$ at radius $r_s$ moving at speed
$\vv_s$ in a field of particles with mass density $\rho_\up{h}(r_s)$ that
move with a Maxwellian distribution of velocities of dispersion $\sigma$
is subject to a dynamical friction acceleration equal to
\beeq
a_\up{df} = - {16\sqrt{\pi} \ln\Lambda G^2 M_s \rho_\up{h}(r_s) \over
\vv_s^2} 
\left[{\sqrt\pi\over4}{\rm erf}~(X)-{X\over2}e^{-X^2}\right] ~,
\label{eq:dfe}
\eneq
where $X=\vv_s/\sqrt{2}\sigma$. Note that the exact value of $a_\up{df}$
in a Jaffe sphere is different because the velocity distribution is
non-Gaussian and is described instead by Dawson's integrals \citep{jaf,bin},
but we neglect this small difference for
simplicity. The Coulomb logarithm term is $\Lambda=b_\up{max}/b_\up{min}$,
where $b_\up{max}$ is taken as the radius $r_s$, and $b_\up{min}$ is the
larger of the satellite radius $R_s$ and $GM_s/\vv_s^2$ (roughly
the impact parameter at which the deflection angle would be 1 radian for
a point mass).

This dynamical friction acceleration results in a loss of orbital
energy $d\epsilon/dt = - \vv_s a_\up{df}$. Assuming the satellite moves on a
circular orbit $r_c$ with $\vv_c(r_c)$ given $\epsilon_0$,
and using $d\epsilon/dr_c=\vv_c\, d\vv_c/dr_c + \vv_c^2/r_c$, we
find that the time derivative of the orbital radius, $\dot r_c$, is
\beeq
{\dot r_c \over r_c} \left(1 + {d\ln \vv_c \over d\ln r_c} \right) =
- {a_\up{df}\over \vv_c} ~,
\eneq
which yields, after using equation (\ref{eq:vcjaffe}),
\beeq
{\dot r_c\over r_c} = {-16\sqrt\pi\ln\Lambda \over P_{\rm orb}}
{M_s\over \MH}{{\RH(r_c+\RH)}\over{r_c(r_c+2\RH)}}
\left[{\sqrt\pi\over4}{\rm erf}~(X)-{X\over2}e^{-X^2}\right] ,
\label{eq:dfe}
\eneq
where $P_\up{orb}=2\pi r_c/\vv_c$ is the orbital period of the satellite.
We use this equation in conjunction with formula (\ref{eq:sig}) for the
velocity dispersion at radius $r_c$ to evaluate the rate at which the
orbit of every satellite decreases due to dynamical friction. For a more
accurate calculation, this rate should be computed as a function of
$r_s$ and the orbital eccentricity, but here we apply a constant rate of
orbital energy decrease, independent of the eccentricity. 
We also require that the relative rate of change of the angular momentum
per unit mass, $L=r_c \vv_c$, is independent of eccentricity:
\beeq
{\dot L \over L} = {\dot r_c \over r_c} + {\dot \vv_c \over \vv_c} =
{\dot r_c \over r_c}\, {r_c+2\RH\over 2(r_c+\RH)} ~.
\label{eq:dfa}
\eneq

\subsection{Adiabatic Orbital Variations}

  As a halo increases its mass accreting new matter, the orbit of a
satellite will change in response to the time variation of the
potential. We use adiabatic invariance to calculate the orbital
changes due to the slow, gradual increase in the mass interior to the
satellite orbit. Conservation of the action variable implies that the
product $r_c \vv_c$ must be conserved. Using the expression
(\ref{eq:vcjaffe}) for $\vv_c(r_c)$, we find that the orbital radius
must vary at a rate
\beeq
\left({r_c+2\RH\over r_c+\RH}\right){\dot r_c\over r_c}=
{\dot \RH\over r_c+\RH}-{\dot \MH \over \MH} ~,
\eneq
where $\dot\MH$ and $\dot \RH$ are the time derivatives of the mass and
radius of the halo. We also require that the angular momentum of the
orbit is conserved to determine the eccentricity variation.

\subsection{Tidal Mass Loss}

  Subhalos are subject to mass loss due to tidal gravitational forces
from their host halo (e.g., \citealt{bin,gne1}). It is useful to 
define the tidal radius $r_\up{t}$, the radius of the subhalo beyond
which the tidal force from the host halo is stronger than the internal
gravity of the subhalo. Particles outside $r_\up{t}$ typically become
unbound, resulting in mass loss.

  To model the tidal mass loss, we follow the simple prescription that
each time a subhalo passes by its pericenter it loses all the mass that
is outside the tidal radius $r_\up{t}$. This tidal radius is obtained by
requiring that the average density within $r_\up{t}$ in the subhalo and
the average density of the host halo within the subhalo orbit are
roughly the same:
\beeq
r_\up{t}=\left[{M_s\over\alpha \MH(<D)}\right]^{1/3}D ~,
\eneq
where $D$ is the distance to the subhalo from the host halo center, and
the dimensionless parameter is $\alpha=2$ for the Roche limit and
$\alpha=3$ for the Jacobi limit. \citet{hay} investigated the tidal mass
loss of dark matter subhalos in a static potential using $N$-body
simulations and found that the Jacobi limit is a good approximation.
Here we adopt the Jacobi limit ($\alpha=3$).

  The mass and radius of the subhalo after each passage by its
pericenter are modified according to
\beeq
M'_\up{h}=\MH(<r_\up{t})=\MH{r_\up{t}\over r_\up{t}+\RH},~~~R'_\up{h}=\RH
{r_\up{t}\over r_\up{t}+\RH} ~.
\label{eq:tid}
\eneq
This ensures that the velocity dispersion of the subhalo at small radius
remains unaffected.

\subsection{Random Orbital Deflections}

  Subhalos have their orbits perturbed by the gravitational interactions
among themselves. These random perturbations may reduce the orbital
pericenter, thereby hastening the destruction of the satellite, or they
may also increase the pericenter and allow the satellite to survive.
Here we adopt a simple prescription to account for these orbital
perturbations, based on the fact that,
to conserve the total energy of the system,
the energy lost by satellites
owing to the dynamical friction process must equal the energy gained
by all the matter in the host halo (both smooth matter and satellites)
owing to random encounters. We first compute the contribution to orbital
perturbations due to the dynamical friction of each $i$-th subhalo. The
orbital energy lost over a short time interval $\Delta t$ is
\beeq
\Delta E_i={\dot E_i}\Delta t=F_\up{dyn}\vv_c\Delta t.
\eneq
We note that the contribution from smoothly distributed mass of the
host halo is negligible compared to the contribution from subhalos.
Then, using the idea that the energy $\Delta E_i$ is transferred to all
the mass, $\Delta M_i$, of the host halo between the radii in which the
orbit of the $i$-th subhalo is comprised, the orbital perturbation
$\Delta\vv_j$ applied to the $j$-th subhalo is obtained by summing over
all the contributions from the other subhalos,
\begin{eqnarray}
{1\over2}\Delta\vv_j^2&=&\sum_{i\neq j}w_{ij}{\Delta E_i\over\Delta M_i}=
\sum_{i\neq j}w_{ij}\vv_{c_i}^2\left({\RH\over r_{c_i}}\right)
\left({M_{s_i}^2\over \MH\Delta M_i}\right) \nonumber \\
&&\times\left({\vv_{c_i}\Delta t\over r_{c_i}}
\right){4\over\sqrt{\pi}}\ln\Lambda\left[{\sqrt{\pi}\over4}{\rm erf}(X_i)
-{X_i\over2}e^{-X_i^2}\right],
\label{eq:pert}
\end{eqnarray}
where $\Delta M_i\equiv \MH(<r_{A,i})-\MH(<r_{P,i})$ is the mass of the
host halo between the radii $r_{A,i}$ and $r_{P,i}$, which are the
apocenter and pericenter radii of the orbit of the $i$-th subhalo.
The summation is made only over the subhalos whose orbits overlap with
the orbit of $j$-th subhalo, and we adopt a simple weighting factor,
\beeq
w_{ij}={\min(r_{A,i},r_{A,j})-\max(r_{P,i},r_{P,j})\over r_{A,j}-r_{P,j}},
\eneq
to account for the fraction of the time that the $j$-th halo spends in
the region where it can interact with the $i$-th halo. Although this
time fraction is not linear with the radial width of the overlap region,
we use this
form for simplicity. Computing exactly the fraction
of the energy that each satellite deposits in each radial shell would
not introduce large differences on the statistical results in any case.

  In summary, at each interval $\Delta t$, we add an orbital
perturbation with amplitude $\Delta\vv_j$ and arbitrary direction to the
velocity of each subhalo, deflecting its orbit.
We note that this
method of perturbing the orbits of the satellites yields results that
are independent of the timestep $\Delta t$ in the limit of a small
timestep, as they should be. Because random perturbations in the
velocity are added quadratically, the amplitude of the velocity
perturbation at each timestep should indeed be proportional to
$\sqrt{\Delta t}$, as obtained in equation (\ref{eq:pert}). 

\subsection{Transfer of Subhalos}
\label{sec:de}

  When a subhalo is being tidally disrupted, smaller satellites within
the subhalo
may be moved out of the system by the tidal forces, and
have their orbit transferred to the larger halo within which the subhalo
is orbiting. To incorporate this effect in our simple model, we remove
satellites from their parent subhalos whenever their orbital radius
is greater than the tidal radius $r_\up{t}$,
and we assign to them a new orbit in the larger halo. Since escapees
follow the motion of their original subhalo, we assume that the new
satellite orbit is close to that of their parent in the larger halo,
although perturbed randomly by a velocity perturbation
$\Delta\bdv{\vv}$, which is applied at 
the position of the parent in the larger halo.
The velocity perturbation given to an escaping ``sub-subhalo''
is obtained by computing the velocity dispersion
$\sigma(r)$ at its position in its original subhalo
and generating three one-dimensional velocity perturbations along
each axis from a Gaussian distribution with this dispersion.

  It happens occasionally that one of the satellites of a subhalo
acquires the escape velocity $\vv_\up{esc}^2(r) =
(2G\MH/\RH)\ln(1+\RH/r)$ owing to the random orbital perturbations.
Then, it simply escapes the subhalo and is placed in a new
orbit within the larger halo which is the same as that of its parent
halo; in this case 
we add a random velocity perturbation with magnitude
$\Delta\vv$ computed from energy 
conservation: $\Delta\vv^2=\vv^2(r)-\vv_\up{esc}^2(r)$.
When this happens to a satellite of a parent halo that does not belong
to any larger halo, the escaping subhalo is simply removed (in
practice we find this does not occur very frequently).

\begin{figure}[t]
\centerline{\epsfxsize=3.5truein\epsffile{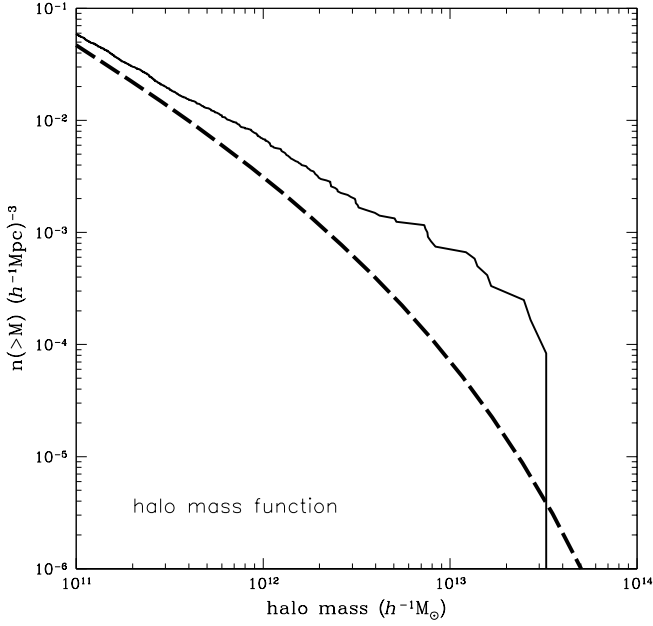}}
\caption{Halo mass function at $z=2$. {\it Solid curve:}
Halo mass function at $z=2$ in a high-density region
that forms a halo of mass $\MH=10^{15}\hmsun$ at $z=0$.
{\it Dashed curve:} Globally averaged halo mass function at $z=2$. }
\label{fig:imf}
\end{figure}

\begin{figure*}[t]
\centerline{\epsfxsize=5.5truein\epsffile{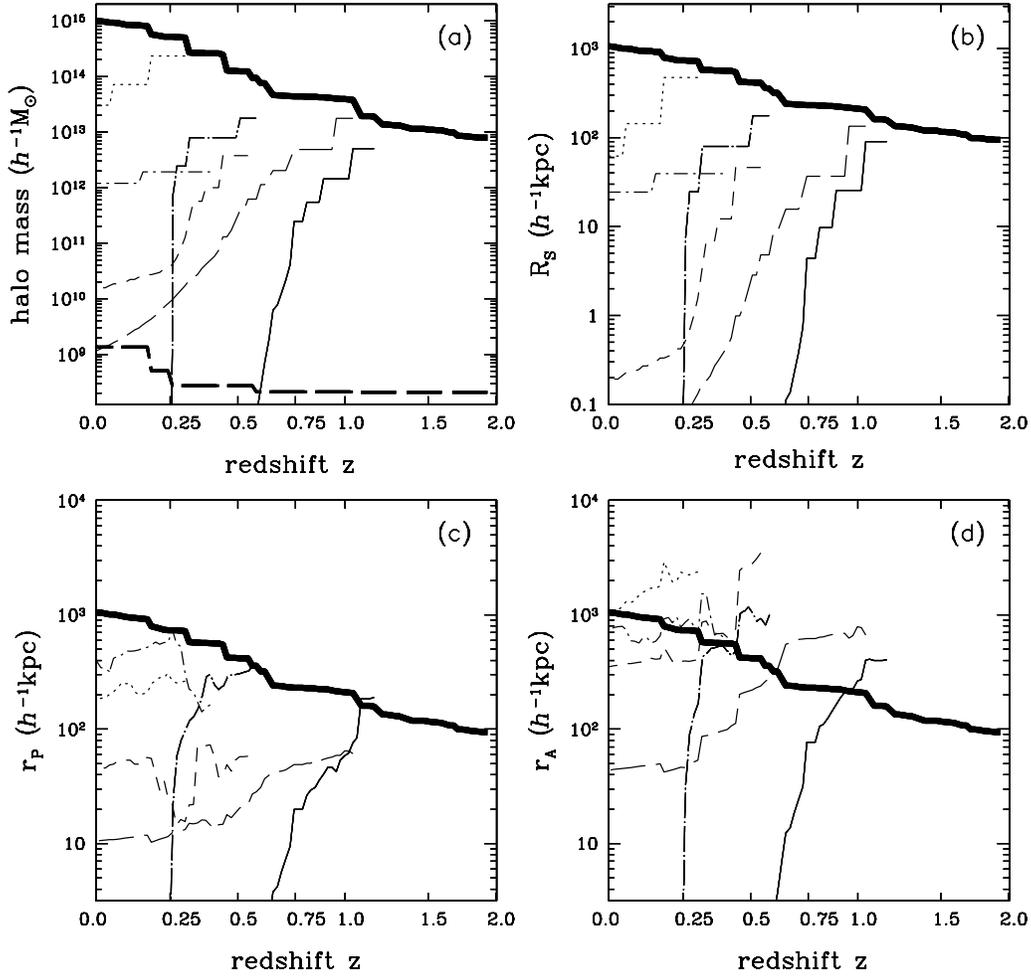}}
\caption{Examples for the dynamical evolution of six subhalos hosting a
black hole, after they merge into the large cluster halo that reaches
a mass $M=10^{15}\hmsun$ at $z=0$. The thin lines in the four panels
show the subhalo mass, radius ($R_S$), pericenter ($r_P$) and apocenter
$r_A$) of the six satellites as a function of redshift. In the upper left
panel, the thick curves are the parent halo ($solid$) and black hole
($dashed$) mass. The thick solid curve in the other three panels is the
luster radius ($R_\up{h}$).}
\label{fig:evo}
\end{figure*}

\begin{figure}[t]
\centerline{\epsfxsize=3.5truein\epsffile{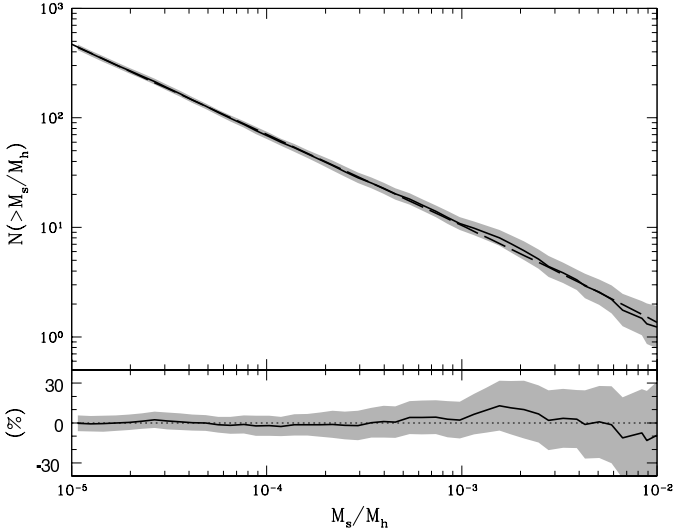}}
\caption{ {\it Solid line:} Number of subhalos found within the $10^{15}
\hmsun$ halo at $z=0$ with a mass fraction above a value $M_s/M_h$,
obtained in our model from the average of 10 numerical realizations.
{\it Shaded region:} statistical uncertainty, computed
from the error on the mean of the ten realizations.
{\it Dashed line:} Analytic subhalo mass function found to fit
results of full numerical simulations of halo substructure
\citep{vale}. 
{\it Bottom panel:} Percentage deviation of our calculation
relative to this fit.}
\label{fig:shmf}
\end{figure}

\begin{figure*}[t]
\centerline{\epsfxsize=5.5truein\epsffile{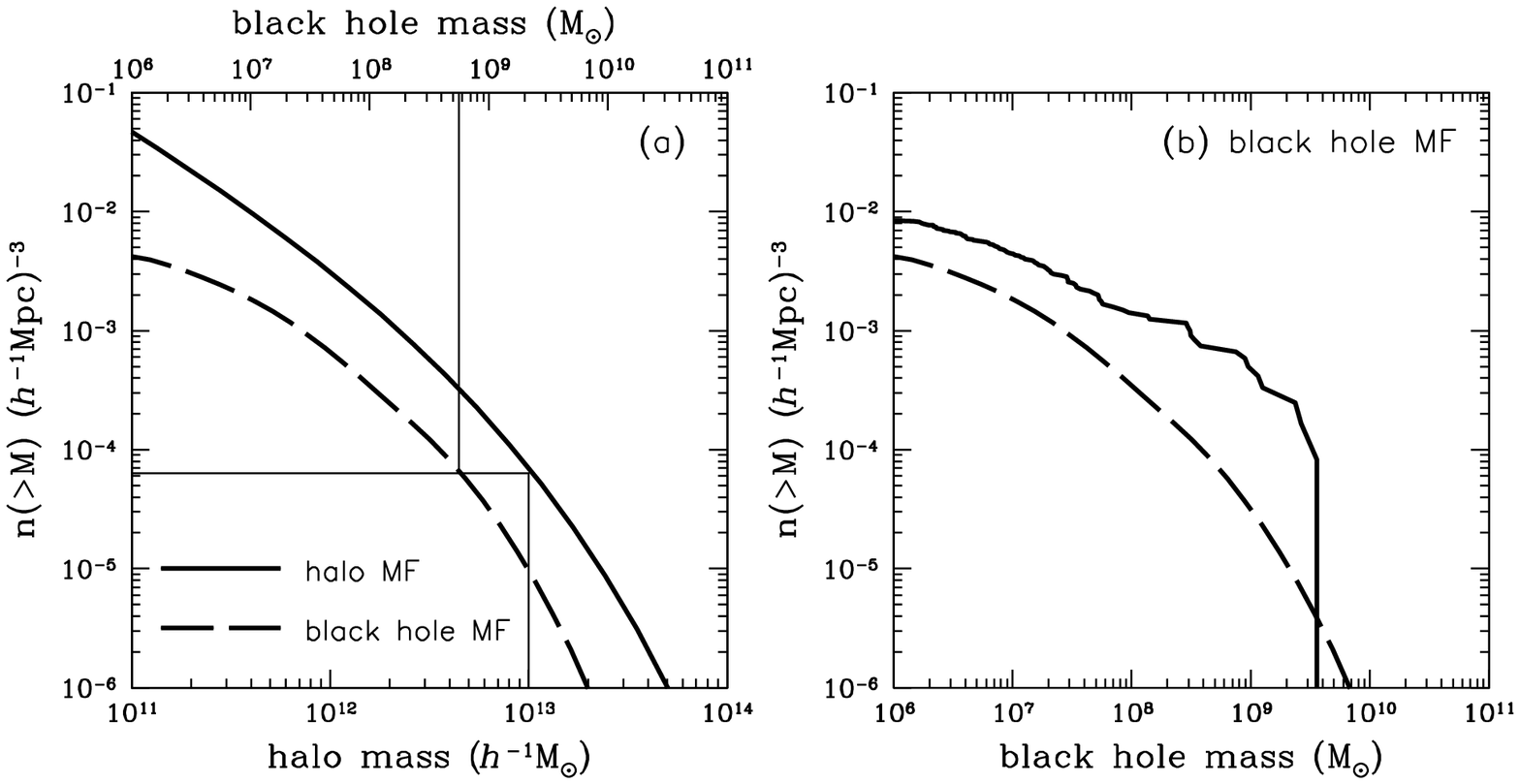}}
\caption{Populating halos with black holes at $z=2$. $(a)$ Assumed
relation of the halo and black hole masses, determined from their
cumulative global mass functions at $z=2$. As an example, black holes of mass
$\sim6\times10^8\msun$ are placed in halos of mass $\sim10^{13}\hmsun$
by matching their cumulative number densities. $(b)$ {\it Solid curve:}
black hole mass function in the cluster region at $z=2$;
{\it dashed curve:} global black hole mass function at $z=2$.}
\label{fig:imf2}
\end{figure*}

\section{Numerical Model}
\label{sec:num}

\subsection{Merger Tree}
\label{sec:mt}

  We use the extended Press-Schechter formalism described in
\S~\ref{sec:mh} to generate the mass accretion histories of halos.
We adopt a flat $\Lambda$CDM cosmology where the present 
Hubble constant is $h\equiv H_0/(100~\kms{\rm Mpc}^{-1})=0.7$, the matter 
density is $\Omega_m=0.3$, the baryon density is $\Omega_b=0.02h^{-2}$, the
power spectrum normalization is determined by $\sigma_8=0.9$, and the
primordial spectral index is $n=1$ (e.g., \citealt{wmap,max2}). The
power spectrum shape is obtained using the transfer function of
\citet{eis}.

  We are particularly interested in this paper in studying the origin
of black holes (and the contribution of mergers to their growth) in 
massive galaxy clusters, such as the Coma cluster. We therefore
generate a merger tree starting at $z=0$ with a halo mass
$\MH=10^{15}\hmsun$, and identify all the progenitor halos with mass
larger than the mass resolution $\Mres$.
In order to accurately follow
the growth and accretion history of halos with masses near $\Mres$, we
actually generate the merger tree down to a smaller mass, which for a
halo of mass $M_{h1}$ is set to the minimum of $10^{-3} M_{h1}$ and
$\Mres$. In this way, the variability of the halo mass growth due to the
stochastic accretion of individual halos is adequately reproduced.
However, only halos
with mass above $\Mres$ are included when we follow their dynamical
evolution and the history of their central black hole. 

  We adopt $\Mres=10^{10}\hmsun$ for the halos for which we follow the
dynamical evolution. As we shall see in \S 5.1, the halos in which we
place black holes in our model are of larger mass 
($\gg 10^{11}\hmsun$) 
than our resolution limit, and the halos with no black holes are
included in the calculation only for the purpose of taking into account
the orbital perturbations they cause on the larger subhalos. We use a
timestep for the merger tree corresponding to $\Delta z = 10^{-4}$,
which is small enough to ensure that $P\ll1$ in equation~(\ref{eq:frag}).
We have verified the validity of our Monte-Carlo procedure by 
generating many merger trees for a large number of halos with the $z=0$
Press-Schechter mass function, and comparing the obtained mass functions
at high redshift to the analytic Press-Schechter mass function.

The halo mass function at $z=2$ for a high-density region that
is constrained to assemble into a halo of mass $\MH=10^{15}\hmsun$ at
$z=0$ is shown in Figure~\ref{fig:imf} (solid line), compared to the
globally averaged halo mass function at $z=2$ (dashed line). Hereafter,
we use the term ``globally averaged'' to refer to a halo mass functions
averaged over the whole universe, without any restrictions to regions
with special properties.
As expected, massive halos are more abundant than
average in a region selected for forming a massive cluster at a later
epoch. The average of ten different realizations of a merger tree was used
for the statistical results presented below in Figures~\ref{fig:shmf}
and \ref{fig:final}.

\subsection{Examples of Dynamical Evolution}

  It is useful to illustrate with a few examples the way that the
dynamical processes described in \S~\ref{sec:dyn} work to determine the
evolution of the subhalos, and ultimately the merger rates of black
holes in our model. Figure~\ref{fig:evo} plots the 
redshift-evolution of the pericenter ($r_p$),
apocenter ($r_A$), Jaffe radius ($R_s$), and mass of six subhalos 
with black holes, which we have selected as
illustrative examples (not randomly) from the merger tree for the
formation of a halo with mass $\MH=10^{15}\hmsun$ at $z=0$. We
generally refer to the largest halo present in the merger tree 
as the ``cluster halo.''

  In our model, a halo grows in mass from accretion and mergers as long
as it merges only with objects smaller than itself, and it becomes a
subhalo when it merges into a larger object. From that point on, the
mass of the subhalo can only decrease as it experiences tidal
disruption. In Figures~\ref{fig:evo}$a$ and \ref{fig:evo}$b$, the masses and radii of the
subhalos are shown starting at the time they merge into the cluster halo.
The decrease in mass and radius occurs at each pericenter passage,
according to equation~(\ref{eq:tid}). The thick solid and long dashed lines
in Figure~\ref{fig:evo}$a$ show the mass of the cluster halo and of the central
black hole, while the radius of the cluster halo is shown as
the thick solid line in the other three panels.

  The evolution of the orbits is shown in Figures~\ref{fig:evo}$c$
and \ref{fig:evo}$d$,
which plot the pericenter and apocenter as a function of time. There
is an average tendency for the orbits to shrink because of dynamical
friction. The rate at which the orbits shrink increases with subhalo
mass and with proximity to the center. As the subhalos approach
the center, their masses and radii are reduced by tidal disruption. It
should be noted here that when the mass of both the subhalo and cluster
halo inside radius $r$ vary linearly with $r$ (as is the case for Jaffe
spheres in the inner parts), the mass of a subhalo after tidal
disruption at pericenter $r_p$ will be proportional to $r_p$, and the
ratio of the subhalo mass to the cluster halo mass enclosed within the
subhalo orbit remains constant. Therefore, the ratio of the dynamical
friction time to the orbital time remains constant also, and the
subhalo continues to spiral toward the center in a constant number of
orbits in each logarithmic radial interval. The subhalo is therefore
completely destroyed in a finite time, and the black holes are then
assumed to merge. In the example in the figures, two halos are
destroyed in the center at $z\simeq 0.56$ and $z \simeq 0.25$, causing
the jumps in the mass of the central black hole. The mass ratios of the
two mergers are 30\% and 83\%, respectively. 
There is one more black hole that merges at $z\simeq 0.16$ with a
mass ratio of 168\% (not shown in the Figure for the sake of clarity).
Another four halos
are shown which are gradually decreasing their orbital radius but still
survive by $z=0$. The halos added to the cluster at $z\simeq 0.25-0.3$
have a very slow rate of orbital decay, and their pericenter is large
enough to avoid any tidal disruption. Their apocenters and pericenters
show random variations owing to the gravitational encounters among
subhalos.

  In general, subhalos are destroyed when the dynamical friction time
is shorter than the age of the system. Subhalos that avoid tidal
disruption have an initial dynamical friction time long compared to the
age of the universe. If the cluster halo were to remain perfectly
static, eventually all subhalos would spiral to the center given a
sufficiently long time. However,
this does not happen because as the cluster halo continues
to grow in mass, large subhalos that are continuously merging cause
random perturbations on the orbits of smaller subhalos which can
increase their orbital radii and therefore their dynamical friction
times. Additionally, whenever the halo within which a satellite is
orbiting merges into a larger system and is tidally disrupted, the
satellite is placed in an orbit in the larger system with a much longer
dynamical friction time. In this way, small subhalos can survive
indefinitely orbiting in larger halos as long as the halos continue to
merge and grow.

  The important dynamical effects determining the rate of black
hole mergers obtained in our model are the fraction of satellites that
can spiral all the way into the center of their parent halo by dynamical
friction before they are scattered, and the rate at which they are
tidally disrupted as they pass near the halo center. Hence, the black
hole merger rates we present in \S 6 are most sensitive to the rate of
dynamical friction and mass loss by tidal disruption as a function of
orbital eccentricity and semimajor axis, as well as the rate of orbital
perturbations by other satellites. 

  As a test of our analytic model, we compare the subhalo mass
function we obtain in the cluster halo at $z=0$, plotted in
Figure~\ref{fig:shmf}, with results that have been obtained from
numerical simulations. We include only subhalos on orbits
with a circular radius smaller than the cluster halo radius, to compare
with numerical simulation results where subhalos are identified as
bound objects inside the virial radius of the parent halo.
Our subhalo mass function is computed from the average of ten
realizations of the merger tree of the cluster halo. We fit the result
to a Schechter mass function for the subhalo mass function,
\beeq
N(M_s|\MH)dM_s=A\left({M_s\over\beta\MH}\right)^{-\alpha}\exp\left(-{M_s\over
\beta\MH}\right){dM_s\over\beta\MH},
\eneq
where $\alpha$ is the power-law slope and $\beta\MH$ is the cutoff mass
\citep{vale}. The normalization is given by
\beeq
 A = { \gamma/\beta \over
\Gamma[2-\alpha]-\Gamma[2-\alpha,M_{s,\up{max}}/\beta\MH]} ~,
\eneq
where the total mass fraction in subhalos is $\gamma$,
$M_{s,\up{max}}=0.5\MH$ is the maximum subhalo mass, and $\Gamma[x]$
and $\Gamma[a,x]$ are the Gamma and the incomplete Gamma functions,
respectively. We fix the cutoff mass fraction to $\beta=0.3$
\citep{shaw} and fit the analytic function to our results, allowing the
parameters $\alpha$ and $\gamma$ to vary. 
The resulting best-fit subhalo mass function is shown as a dashed line,
and has values $\alpha=1.82$ and $\gamma=0.12$, in good agreement with typical
values found in numerical simulations, $\alpha=1.7 - 1.9$ and
$\gamma \simeq 0.1$ \citep{lucia,shaw,zentner}.
Our model basically agrees with the results found in simulations
within the uncertainty.

\section{Black Hole Mass Function in Clusters at $z=2$}
\label{sec:res}

 This section describes the way we initially populate halos with
central black holes. Our dynamical evolution model described previously
tells us the rate at which halos merge all the way to their central
cusps, and we assume that the merger of the central black holes follows
immediately after the merger. The simple model that we now describe for
the initial population of black holes in halos allows us to derive a
rate of black hole mergers from the rate of halo mergers that proceed all
the way to their centers.

  A full model of the evolution of the black hole mass function
including the effects of gas accretion and mergers would include the
gradual growth of black holes from gas accretion as constrained by
observations of the quasar luminosity function at each redshift, and
would simultaneously follow black hole mergers from the dynamical
evolution of host halos. Here we
make a simplifying assumption that separates the two evolutionary
factors for the growth of black holes:
gas accretion and mergers. Observationally, we know that most
of the radiative energy of luminous quasars was emitted over a narrow
range of redshift, $1.5 \lesssim z \lesssim 4$ 
\citep[e.g.,][]{croom,richards}. Hence, we make the approximation
that
gas accretion creates all the black holes instantaneously at $z=2$,
roughly at the epoch of the peak in quasar activity, and that
subsequently the black hole mass function evolves only by mergers. We
choose to place the black holes in halos at $z=2$ with their observed
mass function at $z=0$. If black hole mergers cause relatively small
changes in the black hole mass distribution, then
the evolved mass function should still be approximately correct.

  The black hole mass function at $z=0$ is often estimated by using
observed luminosity functions or velocity dispersion measurements of
galaxies, and observed correlations with the central black hole masses.
Here we use the black hole mass function obtained by \citet{adam}.
This is derived from the distribution of early-type galaxy velocity
dispersions \citep{early} using the relation between the black hole
mass $\MBH$ and velocity dispersion $\sigma$ in \citet{mbh}, and adding
also a contribution from spiral galaxies (these become the dominant host
galaxies of black holes with $\MBH\leq10^8\msun$, using the estimate of
\citet{aller} ). We also include an intrinsic scatter to the
$\MBH-\sigma$ power-law relation of 0.5 dex, with a log-normal
distribution in black hole mass.

  We assign black holes to host halos at $z=2$ by assuming that
all halos that have not merged into any larger object contain a black
hole by $z=2$, and that there is a monotonically increasing, one-to-one
relation between the halo mass, $\MH$, and the
corresponding black hole mass, $\MBH(\MH)$, at this redshift. This
implies that the cumulative number densities of
halos with mass above $\MH$ must be equal to the cumulative number
density of black holes with mass above $\MBH( \MH)$. This is
illustrated in Figure~\ref{fig:imf2}$a$, where the
globally averaged halo and black hole mass functions are plotted.
As an example, halos of mass $\MH = 10^{13}\hmsun$ have the same number
density as black holes of mass $6\times10^8\msun$, so we place a black
hole of this mass in each
halo of mass $\MH$. We apply this relation down to a minimum black hole
mass of $10^6\msun$, which corresponds to a halo mass $\MH \geq
8 \times10^{11}\hmsun$.

  Note that the merger tree and the dynamical evolution of halos are
computed from $z\gg 2$, even though black holes are only assigned to
halos at $z=2$. Furthermore, the black holes have $no$ impact on the
dynamical evolution of their host halos in our model. When orbiting
halos merge all the way to their centers, we simply add the mass of
their two black holes to obtain the mass of the new central black hole.

  In Figure~\ref{fig:imf2}$b$, 
the black hole mass function in the cluster halo at
$z=2$ obtained in this way is compared to the globally averaged black
hole mass function.
  The black hole mass function in the cluster halo that we simulate is
of course not the same as the globally averaged black hole mass
function because, owing to the biased spatial distribution of halos,
the region that collapses into a massive halo at $z=0$ contains an
excess density of halos at $z=2$ and therefore an excess density of
black holes compared to the global average. 
 The mass function is shown as the number of black holes per unit
of comoving original volume before the cluster collapses.
Therefore the larger number density of black holes is not due to the
physical collapse of the cluster halo, and reflects only
the biased distribution of black holes.

\begin{figure*}[t]
\centerline{\epsfxsize=5.5truein\epsffile{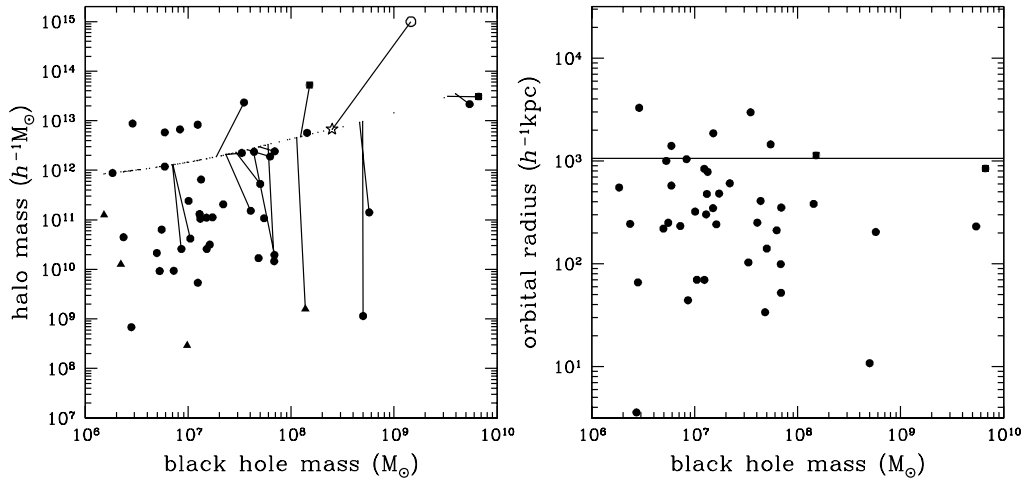}}
\caption{Distribution of satellite halos containing a nuclear
black hole in one realization of the cluster halo.
{\it Left panel}: Filled symbols indicate masses of black holes and
their host subhalos at $z=0$. Small dots give the initial relation
between halo mass and black hole mass at $z=2$. For black holes that
have merged, the initial and final masses are connected; for other
black holes, only the halo masses have changed.
The asterisk
and open circle identify the central halo and black hole in the cluster.
Squares indicate subhalos that have their own satellite halos with black
holes, and these satellites are shown as triangles. {\it Right panel}:
Orbital radius of satellites hosting a 
black hole in the cluster (the central black hole is
not plotted).
The horizontal bar is the radius of the cluster.}
\label{fig:dist}
\end{figure*}

\begin{figure*}[t]
\centerline{\epsfxsize=7truein\epsffile{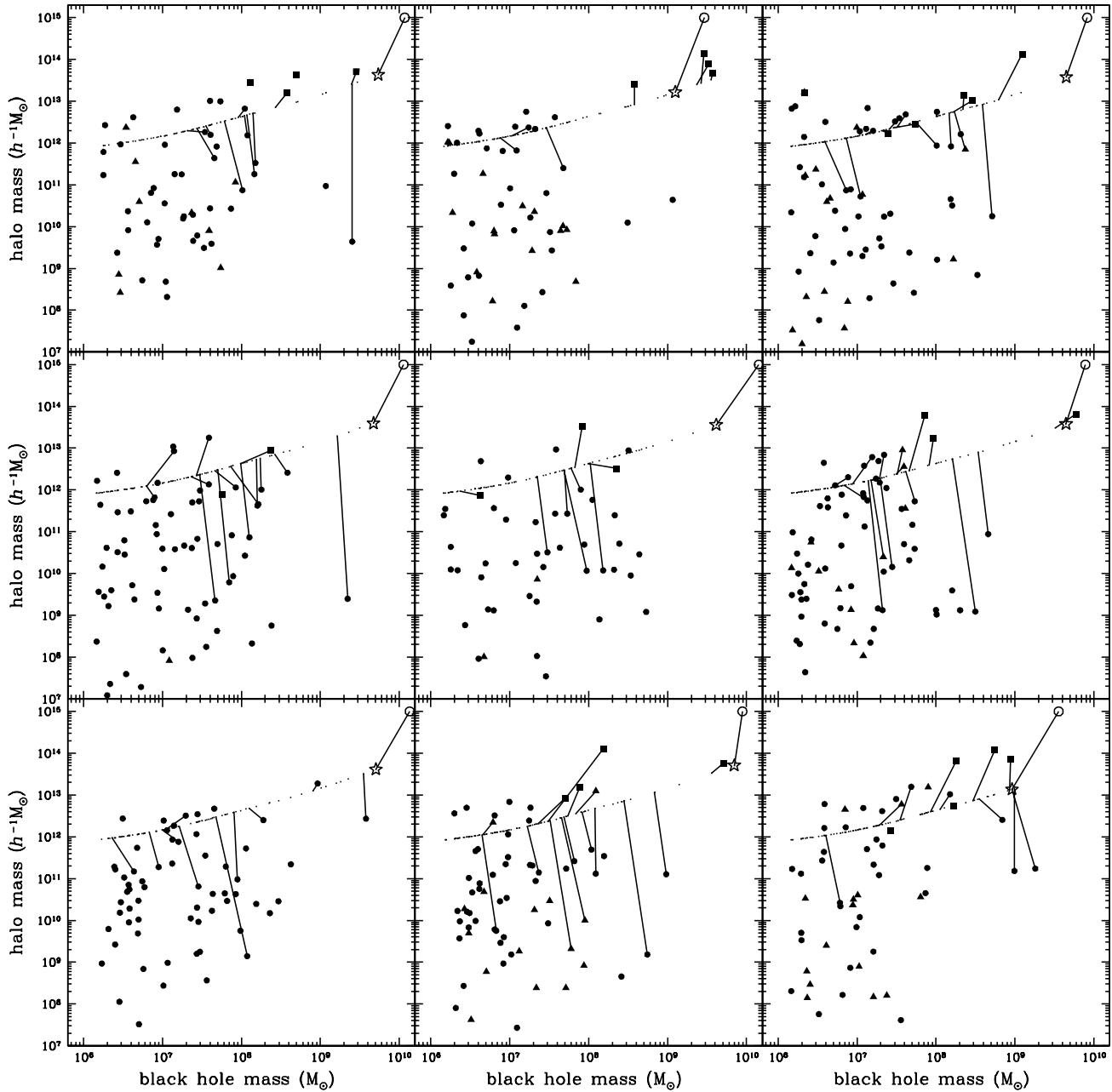}}
\caption{Distribution of the black holes in the nine additional
realizations of the merger tree of the cluster halo.
The symbols are the same as Fig.~\ref{fig:dist}.}
\label{fig:many}
\end{figure*}

\begin{figure*}[t]
\centerline{\epsfxsize=5.5truein\epsffile{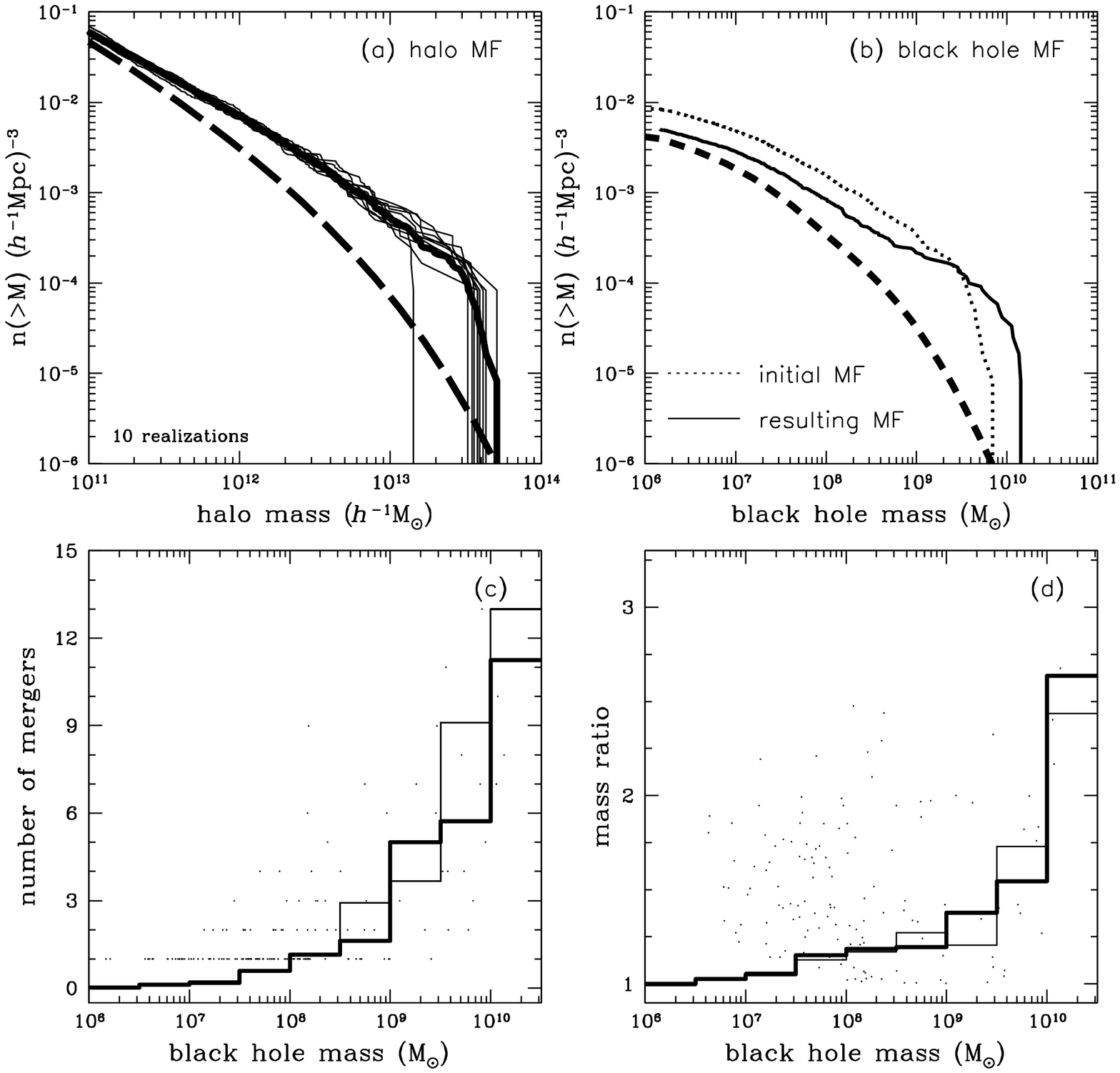}}
\caption{
$(a)$ {\it Thin solid lines:} Halo mass function in the $M=10^{15} 
\hmsun$ cluster at $z=0$ in each realization. 
{\it Thick solid:} Halo mass function averaged over the ten
realizations.
{\it Thick dashed:} Globally averaged halo mass function.
$(b)$ Black hole mass function in the cluster initially (at $z=2$)
and at $z=0$, after the mergers have occurred. The thick dashed
line is the globally averaged black hole mass function.
$(c)$ and $(d)$ Individual small dots are the number of mergers
and mass growth factors for black holes that have merged in ten
realizations. Thick solid lines are averages.
Thin solid lines are the number of mergers and mass ratios,
when interaction between subhalos is ignored.}
\label{fig:final}
\end{figure*}

\section{Impact of Mergers on Black Hole Growth}
\label{sec:impact}

  In this section, we present the results on the impact of mergers on
the black hole mass function in a massive cluster and the way black
holes are distributed in a cluster. We do not consider any
possible ejections of black holes due to gravitational wave recoil, and
we neglect the mass lost due to the emission of gravitational waves
during mergers.

  First, we examine the results of one particular realization of a
$10^{15}\hmsun$ halo. The left panel in Figure~\ref{fig:dist} shows the
initial and final values of the black hole mass and the host subhalo
mass
for all the black holes hosted by satellite halos of the cluster.
The small dots are the values at $z=2$, and they follow the 
black hole mass$-$halo mass relation that we impose as initial
conditions. At $z=0$, the black hole and halo masses are shown as filled
symbols. The central black hole in the cluster is the one hosted by the
main halo with mass $10^{15}\hmsun$, and all other halos have become
satellites in the cluster. Subhalos with their own satellite systems
that have survived
tidal disruption by the main cluster halo are shown as filled squares,
and satellites of subhalos are shown as triangles. Note that the
evolution of the halos from $z=2$ to $z=0$ may either increase the halo
mass due to mergers and accretion of other halos, or decrease the halo
mass due to tidal stripping and disruption when they become subhalos in
a larger halo. On the other hand, the black hole mass can only increase 
if two black holes merge, following a complete disruption of the host
subhalo. To facilitate the identification of progenitor black holes, 
we connect filled symbols ($z=0$) and dots ($z=2$) of the black holes
that have experienced any mergers. The right panel in
Figure~\ref{fig:dist} shows the final orbital radius of each satellite
in the cluster and the black hole mass, showing little correlation
between these two quantities, except perhaps at the highest masses
where the black holes may tend to be hosted by satellites closer to
the cluster center.

  Most of the black holes with masses above $\sim 10^{7.5} \msun$
experience mergers, and some of them increase their mass in a
substantial way. The central black hole is the one that grows the
most, experiencing many mergers and growing its mass by a factor
$\sim 5$. At lower masses, most black holes do not experience any
merger; for these, the filled symbol is at the same black hole mass as
the small dot (note that some small dots do not have any corresponding
large symbol because they have merged with a more massive black hole).
In our dynamical model, the satellites representing the host galaxies of
these black holes form galaxy groups and later become part of a larger
cluster, where their dynamical friction time is too long to produce any
significant decay. Although the outer parts of these satellites may
be tidally disrupted (producing the large reductions in halo mass for
most of the cases shown in the figure), the core of the satellite
containing the black hole survives and remains as a satellite halo.
We need to caution that the rarity of mergers for low-mass black holes
may be caused in our model by the fact that we do not include black
holes with initial mass below $10^6 \msun$, and we place all the black
holes at $z=2$ in field halos (halos that are not satellites in a
larger halo).

  The black hole that ends up in the cluster center is identified as
the one that was present in the original halo which, merging only with
other halos smaller than itself, became the cluster halo at $z=0$. In
any merger-tree simulation of the formation of a cluster, there is one
and only one initial halo that has always merged with other halos smaller
than itself, which is found by tracing back the merger history from the
final cluster and choosing always the most massive progenitor halo at
every merger event. This halo is not necessarily the most massive halo
at $z=2$. Because the growth rates of halos are stochastic, there may be
halos of high mass at $z=2$ which grow slowly, while other halos of
originally lower mass grow faster and become more massive by the time
that all the halos merge to form the final cluster at $z=0$. This is in
fact the case in the example of Figure~\ref{fig:dist}, where the initial
halo that becomes the center of the final cluster is identified by an
asterisk. This central progenitor has a mass of $\sim 10^{12.7}\hmsun$
at $z=2$ and it grows rapidly to become the cluster halo, while several
other halos that were originally more massive (the largest
having nearly $10^{14}\hmsun$ at $z=2$) become satellites as they merge
at late times with the fast-growing halo. When this situation occurs,
the black hole in the central galaxy will not be the most massive one.
In the present example, the black hole in the cluster center would have
a mass of only $10^{9.2} \msun$, while black holes in other satellite
galaxies have masses as high as $\sim 10^{9.8} \msun$.

  This situation, however, is not the most common one, as can be seen
in Figure~\ref{fig:many}, where an additional nine random realizations
of the formation of a $10^{15}\hmsun$ halo are examined. In the majority
of cases, the halo that becomes the final cluster at $z=0$ was the most
massive halo at $z=2$. Out of the ten random realizations we have
performed, there is only one case where black holes in satellite
galaxies are substantially more massive than the black hole ending up at
the center, but there are several where black holes in satellite
galaxies are of comparable mass to the central one. 
We also see that there is
a substantial dispersion for the mass of the central black hole in a
cluster of fixed mass ($10^{15} \hmsun$), going from values as high as
$1.5\times 10^{10} \msun$ to as low as $2\times 10^9 \msun$. If we take the
most massive black hole in the cluster (rather than the central one),
then the low-end of this mass range increases to $4\times 10^{9} \msun$,
still implying a substantial dispersion.

  In practice, the distinction between the black hole in the cluster
center and black holes in satellites is ambiguous observationally.
Clusters typically have substructure and may have more than one galaxy
that could qualify as ``central.'' However, observationally we know
that there is a linear correlation between black hole mass and the
stellar mass of an elliptical galaxy (or the bulge component of a spiral
galaxy). This observed relation tells us that, irrespective of whether a
giant galaxy is considered to be the central one in a cluster or a
satellite, its stellar mass should follow the same correlation with
black hole mass that was imprinted during the quasar epoch at high
redshift. Mergers will increase the mass of a black hole and the
spheroidal stellar component by the same factor (neglecting
gravitational wave emission and ejected stars), so in any clusters
where the central black hole is not the most massive one, it should
probably also be true that the central galaxy is not the one with the
largest stellar mass, and that the most massive black hole is in the
galaxy with the most massive stellar component.

  Figure~\ref{fig:final} summarizes the results of the ten realizations. 
The initial halo mass functions at $z=2$ of the ten realizations are
shown in the first panel, together with the average mass function. The
mass functions in the realizations contain many more massive halos than
average because of the bias introduced when selecting a region that
forms a $10^{15}\hmsun$ halo at $z=0$. The second panel shows the
black hole mass functions. For the same reason, the initial mass
function contains many more massive black holes per unit volume than
the average mass function. Black hole mergers result in a substantial
change in the black hole mass function, in particular a large increase
in the number of black holes of $M \gtrsim 5\times 10^{9}\msun$.

  Because of this increase, our initial conditions in which the $z=2$ black
hole mass function matches the $z=0$ mass function are not fully consistent
with our results. However, one should note that Figure~\ref{fig:final}$b$
shows the change in the black hole mass function in a $10^{15}\hmsun$ cluster,
and that such clusters are rare; the global evolution of the high end of the
mass function will be weaker.

  In the last two panels, thick solid histograms show the number of mergers
experienced by each black hole in terms of the black hole mass, and the
average factor by which the black hole mass increases as a result of
these black hole mergers. Figure 6c shows that most black holes of
$M\gtrsim 10^8\msun$ experience mergers. However, their mass growth is
small except for the most extremely massive black holes, $M \gtrsim
3\times 10^9 \msun$. The majority of black holes, with lower masses,
grow only by a small factor because they often merge with black holes of
much lower mass than themselves.
At low masses, black holes are removed by merging into more massive
objects, causing a decrease in the mass function of a factor of $\sim 2$
(Fig.\ref{fig:final}$b$).
We are cautious in interpreting this result because it may be severely
affected by our approximation of placing the black holes initially at
$z=2$: had we followed the evolution of black holes from a higher
redshift, when many more low-mass halos are present, many more mergers
of low-mass halos (and their central low-mass black holes) would
probably have occurred. The minimum black hole mass of $10^6 \msun$ in
our simulation should probably also be increased in order to correctly
follow the black hole mergers at higher redshift, and a model of black
hole growth from gas accretion that is distributed over redshift
(instead of occurring instantaneously at $z=2$, as we have assumed here)
would need to be included.

A fundamental process affecting the way black hole mergers can occur 
in our model is the mechanism of random orbital perturbations described
in \S~3.5 to take into account the interactions of the satellite halo
containing the black hole with other satellites. If these perturbations
are absent, a satellite can only continue to undergo orbital decay by
dynamical friction, unless its parent halo merges into another object.
The perturbations allow a satellite to randomly gain
orbital energy and in this way avoid tidal destruction. It is therefore
useful to check how much our merger rates are altered if we suppress
these interactions, to see if the results are highly sensitive to their
presence and to the detailed way in which we implement the
perturbations. This is quantified in Figure~7, in panels ($c$) and
($d$), where thin solid histograms show the results with the
perturbations turned off. As expected,
the number of black hole mergers increases, although by a relatively
small amount, while the mass growth remains similar. Hence the
results are not greatly affected by the presence of the perturbations.
We also find that the enhanced disruption of subhalos when perturbations
are turned off results in a
subhalo mass function for which only 7\% of the total mass is contained
in subhalos, less than is found in numerical simulations. 
Thus, our physically motivated recipe for including orbital perturbations
produces better agreement between the analytic model and numerical simulations
(see Fig.\ref{fig:shmf}).

\section{Conclusions}
\label{sec:con}

  We have developed a dynamical model of substructure in dark matter
halos based on numerical merger trees and analytic models of the
dynamical evolution of satellites, with the purpose of modeling the
rates of mergers of nuclear black holes. We have improved on previous
work by tracing the evolution of satellites in a satellite halo, and by
accounting for satellite-satellite interactions. Our model reproduces
the analytic subhalo mass function found previously in numerical
simulations, for $M_s\geq10^{-5}\MH$ in a halo of $\MH=10^{15}\hmsun$.
In this paper, we have focused on examining the effect of mergers on the
abundances of the most massive black holes in  clusters of galaxies,
using a simplified model where black holes have completed their growth
by gas accretion by $z=2$ and remain with a fixed mass after that,
unless they undergo mergers when their host galaxies merge. We have
assumed that the merger of two
satellites in a halo always leads to the merger of their central black
holes (which maximizes the merging rate for black holes).

We find that mergers have an important impact on the most
massive black holes in the cluster. For the most massive black holes in
a cluster, with $M > 10^9 \msun$, a growth in mass by a factor $\sim 2$
is typical. Our ten realizations of a $10^{15}\hmsun$ cluster include four
black holes with a final mass of $1-1.5\times10^{10}\msun$, but without
mergers such black holes would be extremely rare.

  Mergers are less important for black holes of lower mass. The main
effect found in our simulation on low-mass black holes is that they
are depleted by a factor $\sim 2$ because of their mergers into more
massive black holes. This result may be affected by our procedure of
initially populating the halos with black holes at $z=2$, and by the
minimum black hole mass of $10^6 \msun$ that we use.
We also caution that the effect of depletion in the global black hole
mass function may be much weaker than the effect in the cluster
environment, because by the epoch of $z=2$, most of the low-mass black
holes that were formed in a region collapsing in a massive cluster at
present have already merged in larger objects.

  We have also found that the most massive black hole is not always
located in the central galaxy of the cluster, if we define the central
galaxy to be the one that formed in the halo that has always merged with
other halos smaller than itself. In practice, it may be difficult to
assign observationally which galaxy in the cluster corresponds to this
central one; moreover, the linear relation between black hole mass and
bulge stellar mass should not be altered.

  Our model may overestimate the importance of mergers because we have
assumed that, once the nuclei of two satellite halos have approached
each other down to the radii of influence of the black holes, the black
holes will always merge. Actually, two black holes may become stalled
after they have scattered all the stars interior to their orbit as
well as any remaining stars in the loss-cone, and before reaching the
radius of orbital decay by gravitational waves 
\citep{mitchell}, so it may be that 
not all black hole binaries merge. It is
unlikely that a majority of mergers end up in stalled binary
systems, however, because few binary black holes in galactic nuclei
are found, and because the presence of gas is likely to drive the binary
to a merger in any case (Gould \& Rix 2000). Moreover, models that try
to estimate the probability of ejection of black holes during three-body
interactions find that only a small fraction of the black holes can be
lost in this way (Volonteri et al.\ 2003).

  Therefore, the results of our paper support the idea that black hole
mergers are important in shaping the black hole mass function at low
$z$, particularly affecting in a strong way the abundance of the most
massive black holes in rich clusters of galaxies.
In a subsequent paper, we shall examine more generally the impact of
mergers on the evolution of black holes outside of massive clusters.

\acknowledgments
We are grateful to Eliot Quataert for useful
discussion on the last figure. This work was supported by NASA grant
NNG056H776. 
J.~Y. is supported by a Presidential Fellowship from the Graduate School
of The Ohio State University. 
J.~M. is supported by Spanish grants AYA2006-06341 and
AYA2006-15623-C02-01. 
Z. Z. acknowledges the support of NASA through Hubble Fellowship grant
HF-01181.01-A awarded by the Space Telescope Science Institute, which
is operated by the Association of Universities for Research in Astronomy,
Inc., for NASA, under contract NAS 5-26555.

\end{document}